\documentclass[conference]{IEEEtran}
\usepackage{fancyhdr}
\usepackage{amsmath}
\usepackage{amssymb}
\usepackage{graphicx}
\usepackage{booktabs}
\usepackage{amsfonts}
\usepackage{multirow}
\usepackage{algorithm}
\usepackage{algorithmic}
\usepackage{mathrsfs}
\usepackage{bm}
\usepackage{exscale}
\usepackage{relsize}
\usepackage{setspace}
\usepackage{color}
\usepackage{ulem}

\newtheorem{theorem}{\textbf{Theorem}}
\newtheorem{proposition}{\textbf{Proposition}}

\ifCLASSINFOpdf
\else
\fi

\begin{document}

\title{Generalized Spatial Modulation Aided MmWave MIMO with Sub-Connected Hybrid Precoding Scheme}

\author{\IEEEauthorblockN{Longzhuang He, Jintao Wang, and Jian Song}
\IEEEauthorblockA{
Department of Electronic Engineering, Tsinghua University, Beijing, China\\
Tsinghua National Laboratory for Information Science and Technology (TNList), Tsinghua University, Beijing, China\\
Email: helongzhuang@126.com}}

\maketitle
\begin{abstract}
Due to the high cost and low energy efficiency of the dedicated radio frequency (RF) chains, the number of RF chains in a millimeter wave (mmWave) multiple-input multiple-output (MIMO) system is usually limited from a practical point of view. In this case, the maximum number of independent data streams is also restricted by the number of RF chains, which consequently leads to limiting the potentially attainable spatial multiplexing gain. In order to address this issue, in this paper, a novel generalized spatial modulation (GenSM) aided mmWave MIMO system is proposed, which enables the transmission of an extra data stream via the index of the active antennas group and requires no extra RF chain. Moreover, a two-step algorithm is also proposed to optimize the hybrid precoder design with respect to spectral efficiency (SE) maximization. Finally, numerical simulation results demonstrate the superior SE performance achieved by the proposed scheme.


\end{abstract}

\section{Introduction}
The idea of millimeter wave (mmWave) communication has recently attracted substantial research interest due to its potential to significantly boost the throughput for future communication networks \cite{daniels2007emerging}. By incorporating with multiple-input multiple-output (MIMO) techniques, mmWave MIMO is capable of striking a substantial spectral efficiency (SE) gain via precoding at the transmitter \cite{torkildson2011indoor}.

Due to the high cost and low energy efficiency of the radio frequency (RF) chains, the number of RF chains in mmWave MIMOs is usually limited from a practical point of view, which facilitates the application of the well-known \textit{hybrid precoding} \cite{ayach2014spatially}\cite{gao2016energy}. While hybrid precoding achieves a near-optimal performance with a limited number of RF chains \cite{ayach2014spatially}, the maximum number of independent data streams is still restricted by the number of RF chains, which therefore limits the potential spatial multiplexing (SMX) gain. In order to address this issue, the concepts of spatial modulation (SM) and generalized SM (GenSM) have been recently incorporated with mmWave MIMOs in \cite{liu2015SSK}-\cite{ishikawa2016GSM}. In SM/GenSM systems \cite{mesleh2008spatial}-\cite{he2015infinity}, a subset of antennas are randomly activated to transmit classic amplitude-phase modulation (APM) symbols, and the information is thus simultaneously conveyed via the APM symbols and the indices of the active antennas.

However, the previous research on the combination of SM/GenSM and mmWave MIMO, i.e. \cite{liu2015SSK} and \cite{liu2016LOS}, all failed to fully exploit the transmitter's knowledge of the channel state information (CSI) for an optimized precoder design, which leads to severe performance penalty when compared to the optimal MIMO waterfilling capacity. To the best of the authors' knowledge, the design of hybrid precoding in a GenSM-aided mmWave MIMO scenario has not been explored.

In order to address this issue, the major contributions of this paper are summarized as follows.
\begin{enumerate}
  \item A novel GenSM-aided mmWave MIMO scheme with a sub-connected hybrid precoding structure is proposed, where an extra data stream is conveyed via the index of the active antennas group without requiring any extra RF chains.

  \item A closed-form SE lower bound is proposed to quantify the achievable SE of the proposed scheme, which is shown to provide a favorable approximation accuracy with a constant shift.

  \item Using the proposed bound as the optimization target, a two-step algorithm is proposed to iteratively optimize the hybrid precoder in terms of SE maximization. As our proposed scheme constitutes a more generalized hybrid precoding paradigm, substantial SE gain over the conventional precoding schemes is thus achieved, as witnessed by our numerical simulations.
\end{enumerate}

The organization of this paper is summarized as follows. Section \uppercase\expandafter{\romannumeral2} introduces the system model of our proposed GenSM-aided mmWave MIMO. Theoretical SE analysis is provided in Section \uppercase\expandafter{\romannumeral3}. Section \uppercase\expandafter{\romannumeral4} introduces our proposed two-step optimization algorithm. The simulation results are provided in Section \uppercase\expandafter{\romannumeral5}, while Section \uppercase\expandafter{\romannumeral6} concludes this paper.

\textit{Notations}: The lowercase and uppercase boldface letters denote column vectors and matrices respectively. The operators $(\cdot)^T$ and $(\cdot)^H$ denote the transposition and conjugate transposition, respectively. The determinant of a matrix $\mathbf{M}$ is denoted by $\vert \mathbf{M} \vert$, while $\otimes$ represents the Kronecker product.

\section{System Model}
\begin{figure}
\center{\includegraphics[width=0.98\linewidth]{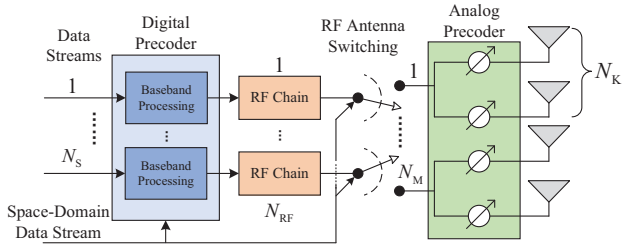}}
\caption{Block diagrams of the proposed GenSM-aided mmWave MIMO scheme.}
\label{Fig_SystemModel}
\end{figure}

We depict our proposed GenSM-aided mmWave MIMO scheme in Fig.\ref{Fig_SystemModel}, where an mmWave MIMO with $N_\text{T}$ transmit antennas (TAs) and $N_\text{R}$ receive antennas (RAs) are considered. As depicted in the figure, the digital precoder, which consists of $N_\text{S}$ baseband processing units, essentially plays the role of conducting \textit{power allocation} to the $N_\text{S}$ input data streams. Note that the power allocation vector simultaneously depends on the input space-domain data stream as well as the instantaneous CSI. This space-information-guided digital precoding is practically achievable due to the high-speed and low-latency advantages of based digital processing.

Moreover, the $N_\text{T}$ TAs are grouped into $N_\text{M}$ antenna groups (AGs), each of which consists of $N_\text{K}$ TAs with $N_\text{T} = N_\text{M} N_\text{K}$. It is required that $N_\text{M} \ge N_\text{RF}$, and the space-domain information randomly assigns the RF-domain symbols to $N_\text{RF}$ out of the $N_\text{M}$ AGs. Finally, similar to \cite{gao2016energy}, $N_\text{K}$ phase shifters (PSs) are equipped in each AG to perform analog precoding. Note that the conventional sub-connected hybrid precoding scheme \cite{gao2016energy} can be conceived as a special case of the proposed framework, when $N_\text{M} = N_\text{RF}$.

The input space-domain information plays the role of specifying an AG activation pattern, which corresponds to a combination of active AGs (AGC). Therefore the total number of legitimate AGCs are given by $M = 2^{\lfloor \log_2 \binom{N_\text{M}}{N_\text{RF}}\rfloor}$ \cite{wang2012generalised}, where $\lfloor \cdot \rfloor$ represents the floor operation, and $\binom{\cdot}{\cdot}$ is the binomial coefficient. By denoting the AGs' indices activated by the $m$-th AGC ($1 \le m \le M$) as $\mathbf{u}_m \triangleq [u_{m1}, u_{m2}, \ldots, u_{m N_\text{RF}} ]^T$ with $1 \le u_{mn} \le N_\text{M}$, the $m$-th \textit{AG-selection matrix} $\mathbf{C}_m \in \mathbb{R}_{N_\text{T} \times N_\text{RF}}$ can therefore be given as follows ($1 \le m \le M$):
\begin{equation}
  \mathbf{C}_m \triangleq \left[\mathbf{e}_{u_{m1}}, \mathbf{e}_{u_{m2}}, \ldots, \mathbf{e}_{u_{mN_\text{RF}}}\right] \otimes \mathbf{1}_{N_\text{K}},
\end{equation}
where $\mathbf{1}_{N_\text{K}} \in \mathbb{R}_{N_\text{K} \times 1}$ is an $N_\text{K}$-dimensional all-one vector and $\mathbf{e}_m$ represents the $m$-th column of $\mathbf{I}_{N_\text{M}}$.

The digital precoder is composed of $M$ diagonal real-valued matrices, i.e. $\mathcal{D} \triangleq \left\{\mathbf{D}_1, \ldots, \mathbf{D}_M\right\}$, where $\mathbf{D}_m = \text{diag}\left(d_{m1}, d_{m2}, \ldots, d_{mN_\text{S}}\right) \in \mathbb{R}_{N_\text{S} \times N_\text{S}}$ is applied, when the $m$-th AGC is selected. Finally, similar to \cite{gao2016energy}, the analog precoder $\mathbf{A} \in \mathbb{C}_{N_\text{T} \times N_\text{T}}$ is given by:
\begin{equation}
  \mathbf{A} \triangleq \displaystyle \text{diag}\left( \frac{1}{\sqrt{N_\text{K}}} e^{j\theta_1}, \frac{1}{\sqrt{N_\text{K}}} e^{j\theta_2}, \ldots, \frac{1}{\sqrt{N_\text{K}}} e^{j\theta_{N_\text{T}}}  \right),
  \label{AForm}
\end{equation}
with $\theta_{n} \in [0, 2\pi)$ denoting the rotation angle of the $n$-th TA.

Therefore the received signal vector $\mathbf{y} \in \mathbb{C}_{N_\text{R} \times 1}$ can be formulated as follows, when the $m$-th AGC is selected:
\begin{equation}
  \mathbf{y} = \displaystyle \sqrt{\rho} \mathbf{H} \mathbf{A} \mathbf{C}_m \mathbf{D}_m \mathbf{x} + \mathbf{n},
  \label{ReceivedSignal}
\end{equation}
where a narrowband MIMO channel matrix $\mathbf{H} \in \mathbb{C}_{N_\text{R} \times N_\text{T}}$ is considered, and $\mathbf{x} \sim \mathcal{CN}(0, \frac{1}{N_\text{S}} \mathbf{I}_{N_\text{S}})$ represents the Gaussian-distributed input signal vector. The average transmit power is given by $\rho > 0$, while $\mathbf{n} \sim \mathcal{CN}(\mathbf{0}, \sigma_\text{N}^2 \mathbf{I}_{N_\text{R}})$ represents the additive white Gaussian noise (AWGN). Besides, we also assume that the digital precoder satisfies the following power constraint:
\begin{equation}
    \displaystyle \sum_{m=1}^M \text{Tr}\left(\mathbf{D}_m \mathbf{D}_m^H \right) \le M N_\text{S},
    \label{FConstraint}
\end{equation}
which leads to a signal-to-noise ratio (SNR) of $\rho / \sigma_\text{N}^2$ at the receiver.

In this paper we adopt the classic clustered Saleh-Valenzuela geometric mmWave channel model \cite{ayach2014spatially}, which is given as:
\begin{equation}
\arraycolsep=1.0pt\def\arraystretch{1.3}
  \begin{array}{lcl}
    \mathbf{H} &=& \displaystyle \gamma \sum_{p=1}^{N_\text{cl}} \sum_{q=1}^{N_\text{ray}} \alpha_{pq} \Lambda_\text{t}(\phi_{pq}^\text{t}, \theta_{pq}^\text{t}) \Lambda_\text{r}(\phi_{pq}^\text{r}, \theta_{pq}^\text{r}) \text{...} \\
    && \displaystyle \mathbf{b}_\text{t}(\phi_{pq}^\text{t}, \theta_{pq}^\text{t}) \mathbf{b}_\text{r}(\phi_{pq}^\text{r}, \theta_{pq}^\text{r}),
  \end{array}
  \label{SVChannel}
\end{equation}
where $\gamma > 0$ is the normalizing factor ensuring $E\{\|\mathbf{H}\|_F^2\} = N_\text{R} N_\text{T}$. For the purpose of brevity, we refer the interested readers to \cite{ayach2014spatially} for more details on the specific channel parameters. In this paper, the parameters in (\ref{SVChannel}) will be specified in accordance to \cite{ayach2014spatially}, unless mentioned otherwise.

\section{Achievable Spectral Efficiency Analysis}
\subsection{Mutual Information Analysis}
The achievable SE of the proposed system is quantified by the mutual information (MI) between $\mathbf{y}$, $\mathbf{x}$ and $m$, i.e.
\begin{equation}
  R(\mathbf{H}, \mathcal{D}, \mathbf{A}) = I(\mathbf{y}; \mathbf{x}, m).
  \label{MI0}
\end{equation}

However, the MI term in (\ref{MI0}) cannot be expressed in a \textit{closed form} due to the discrete channel input $m \in \{1, 2, \ldots, M\}$. In order to provide a computationally efficient approach to quantify (\ref{MI0}), we propose Theorem \ref{theorem1} to provide a closed-form expression $R_\text{LB}(\mathbf{H}, \mathcal{D}, \mathbf{A})$ for lower-bounding $R(\mathbf{H}, \mathcal{D}, \mathbf{A})$:
\begin{theorem}
  The MI term in (\ref{MI0}) can be lower-bounded by the following expression:
  \begin{equation}
  \arraycolsep=1.0pt\def\arraystretch{1.8}
  \begin{array}{lcl}
    R_\text{LB}(\mathbf{H}, \mathcal{D}, \mathbf{A}) &=& \displaystyle \log_2 \frac{M}{(e \sigma_\text{N}^2)^{N_\text{R}}} \text{...} \\
    && \displaystyle - \frac{1}{M} \sum_{n=1}^M \log_2 \sum_{t=1}^M\left|\bm\Sigma_n + \bm\Sigma_t\right|^{-1},
  \end{array}
  \label{MILB}
  \end{equation}
  where $\bm\Sigma_n \in \mathbb{C}_{N_\text{R} \times N_\text{R}}$ is given as ($n = 1, 2, \ldots, M$):
  \begin{equation}
    \bm\Sigma_n \triangleq \sigma_\text{N}^2\mathbf{I}_{N_\text{R}} + \frac{\rho}{N_\text{S}} \mathbf{HA}\mathbf{C}_n \mathbf{D}_n \mathbf{D}_n^H \mathbf{C}_n^H \mathbf{A}^H \mathbf{H}^H.
    \label{SigmanDef}
  \end{equation}
  \label{theorem1}
\end{theorem}

\begin{IEEEproof}
  See the Appendix A of \cite{he2017spatial}.
\end{IEEEproof}

\subsection{Bound Tightness}
We now seek to demonstrate the tightness of the proposed lower bound in Theorem \ref{theorem1}. More specifically, we found that a constant gap exists between the lower bound and the true SE value, when an asymptotically high (low) SNR value is imposed, which is verified by the following proposition.
\begin{proposition}
  With an asymptotically high or low SNR value, a constant gap of $N_\text{R}(1 - \log_2 e)$ exists between the closed-form lower bound $R_\text{LB}(\mathbf{H}, \mathcal{D}, \mathbf{A})$ and the true SE expression $R(\mathbf{H}, \mathcal{D}, \mathbf{A})$.
  \label{proposition1}
\end{proposition}

\begin{IEEEproof}
  See the Appendix B of \cite{he2017spatial}.
\end{IEEEproof}

\begin{table}
\small
\caption{Simulation Parameters}
\newcommand{\tabincell}[2]{\begin{tabular}{@{}#1@{}}#2\end{tabular}}
\centering
\renewcommand\arraystretch{1.2}
\begin{tabular}{l|l|l}
\hline\hline
Symbols                 & Specifications & Typical Values \\\hline
$N_\text{T}$            & Number of TAs                                             & $8$ \\\hline
$N_\text{R}$            & Number of RAs                                             & $8$ \\\hline
$N_\text{K}$            & \tabincell{l}{Number of TAs in \\each antenna group}                       & $2$ \\\hline
$N_\text{M}$            & Number of antenna groups                                  & $4$ \\\hline
$N_\text{RF}$           & Number of RF chains                                       & $2$ \\\hline
$N_\text{S}$            & \tabincell{l}{Number of APM-domain \\data streams}                         & $N_\text{RF}$ \\\hline
$\lambda$               & Carrier's wavelength                                      & $5$ mm \\\hline
\hline
\end{tabular}
\label{TABLESimu}
\end{table}

By adding this constant gap, $R_\text{LB}(\mathbf{H}, \mathcal{D}, \mathbf{A})$ is expected to provide a more accurate approximation to $R(\mathbf{H}, \mathcal{D}, \mathbf{A})$. In order to provide numerical demonstrations, we summarize the typical simulation parameters in Table \ref{TABLESimu}. Besides, as the hybrid precoder design has not yet been discussed, we tentatively assume that $\mathcal{D} = \{\mathbf{I}_{N_\text{S}}, \ldots, \mathbf{I}_{N_\text{S}}\}$ and $\mathbf{A} = \mathbf{I}_{N_\text{T}} / \sqrt{N_\text{K}}$.

\begin{figure}
\center{\includegraphics[width=0.90\linewidth]{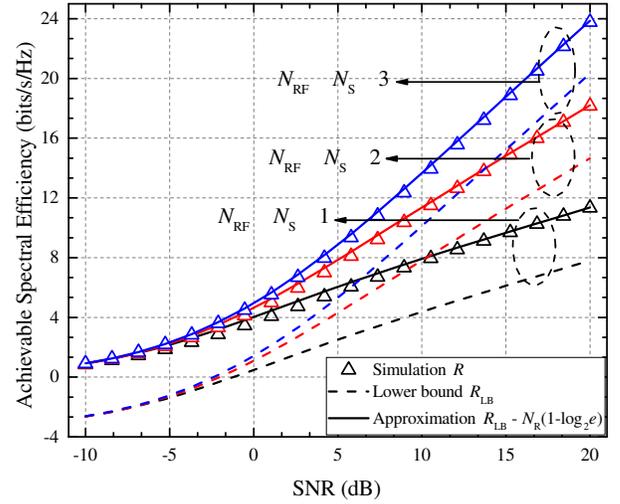}}
\caption{Average achievable SE and SE bounds with various $(N_\text{S}, N_\text{RF})$. The simulation value $R$ and the closed-form lower bound $R_\text{LB}$ are given according to (\ref{MI0}) and (\ref{MILB}), respectively. The approximation is given by $\left[R_\text{LB} - N_\text{R}(1 - \log_2 e)\right]$.}
\label{Fig_BoundTightness}
\end{figure}

We depict the true SE expression $R(\mathbf{H}, \mathcal{D}, \mathbf{A})$ and the SE lower bound $R_\text{LB}(\mathbf{H}, \mathcal{D}, \mathbf{A})$ in Fig.\ref{Fig_BoundTightness}. As it can be seen from the figure, aided with the constant shift, the proposed lower bound $R_\text{LB}$ actually provides a favorable approximation accuracy (the symbols and solid lines are observed to be very close to each other). Since this constant shift is independent from the specific hybrid precoder, applying it thus imposes no impact on the hybrid precoder design in terms of SE maximization, which motivates us to employ $R_\text{LB}$ as the optimization target for the precoder design.

\section{Proposed Algorithm For Precoder Design}
Since $R_\text{LB}$ is capable of providing an accurate approximation to the true SE value, we therefore seek to solve the following optimization (P1) to design the hybrid precoder:
\begin{equation}
\arraycolsep=1.0pt\def\arraystretch{1.3}
  \begin{array}{lcl}
  \text{(P1):    }  & \displaystyle \max_{\mathcal{D}, \mathbf{A}}  & \displaystyle R_\text{LB}(\mathbf{H}, \mathcal{D}, \mathbf{A}) \\
                    & \displaystyle \text{s.t.}                     & \mathcal{D} = \left\{\mathbf{D}_1, \mathbf{D}_2, \ldots, \mathbf{D}_M\right\}, \\
                    &                                               & \displaystyle \sum_{m=1}^M \text{Tr}(\mathbf{D}_m \mathbf{D}_m^H) \le M N_\text{S}, \,\, \text{and} \,\, \mathbf{A} \in \mathscr{A},
  \end{array}
  \label{P1}
\end{equation}
where $\mathscr{A}$ is the feasible set of $\mathbf{A}$ satisfying the definition in (\ref{AForm}). A simultaneous optimization of the precoders $\mathcal{D}$ and $\mathbf{A}$ could be computationally infeasible. In order to facilitate a computationally efficient solution, similar to \cite{zeng2012linear}, we seek to decompose (P1) and \textit{iteratively} solve the following two sub-problems:
\begin{equation}
\arraycolsep=1.0pt\def\arraystretch{1.3}
  \begin{array}{lcl}
  \text{(P2):    }  & \displaystyle \max_{\mathcal{D}}              & \displaystyle R_\text{LB}(\mathbf{H}, \mathcal{D}, \mathbf{A}) \\
                    & \displaystyle \text{s.t.}                     & \mathcal{D} = \left\{\mathbf{D}_1, \mathbf{D}_2, \ldots, \mathbf{D}_M\right\} \\
                    &                                               & \displaystyle \sum_{m=1}^M \text{Tr}(\mathbf{D}_m \mathbf{D}_m^H) \le M \cdot N_\text{S}, \\
  \text{(P3):    }  & \displaystyle \max_{\mathbf{A}}               & \displaystyle R_\text{LB}(\mathbf{H}, \mathcal{D}, \mathbf{A}) \,\, \text{s.t.} \,\, \mathbf{A} \in \mathscr{A},
  \end{array}
\end{equation}
in which (P2) optimizes the digital precoder $\mathcal{D}$ based on a given $\mathbf{A}$, and (P3) optimizes the analog precoder $\mathbf{A}$ upon an given $\mathcal{D}$.

\begin{figure*}
    \begin{equation}
    \arraycolsep=1.0pt\def\arraystretch{3.0}
    \begin{array}{rcl}
      \displaystyle\nabla_{\bm\lambda_m} R_\text{LB}(\bm\lambda) &=& \displaystyle \frac{\rho \log_2 e}{MN_\text{S}} \sum_{n=1}^M \frac{\left|\bm\Sigma_n + \bm\Sigma_m\right|^{-1} \text{diag}\left[ \mathbf{C}_m^H\mathbf{A}^H \mathbf{H}^H  \left(\bm\Sigma_n + \bm\Sigma_m\right)^{-1} \mathbf{HAC}_m \right]}{\sum_{t=1}^M \left| \bm\Sigma_n + \bm\Sigma_t \right|^{-1}} + \text{...} \\
      && \displaystyle \frac{\rho \log_2 e}{MN_\text{S}} \frac{\sum_{t=1}^M \left| \bm\Sigma_m + \bm\Sigma_t \right|^{-1} \text{diag}\left[\mathbf{C}_m^H \mathbf{A}^H \mathbf{H}^H \left( \bm\Sigma_m + \bm\Sigma_t \right)^{-1} \mathbf{HAC}_m\right] }{\sum_{t=1}^M \left| \bm\Sigma_m + \bm\Sigma_t \right|^{-1}}.
    \end{array}
    \label{RLB_Lambda_m}
    \end{equation}
\hrulefill
\end{figure*}

Before providing our solution to (P2), it is worth pointing out that the cost function $R_\text{LB}$ is non-concave over $\mathbf{D}_m = \text{diag}\left(d_{m1}, d_{m2}, \ldots, d_{mN_\text{S}}\right)$, which can be easily verified via a counterexample with $N_\text{R} = N_\text{T} = N_\text{K} = N_\text{M} = N_\text{RF} = M = 1$. However, the following proposition shows that $R_\text{LB}$ is actually a concave function of $\bm\lambda = [\bm\lambda_1^T, \ldots, \bm\lambda_M^T]^T$ with $\bm\lambda_m \triangleq [d_{m1}^2, d_{m2}^2, \ldots, d_{m N_\text{S}}^2]^T$.

\begin{proposition}
The closed-form lower bound $R_\text{LB}$ is a concave function of the \textit{power allocation vector} $\bm\lambda$.
\label{proposition2}
\end{proposition}

\begin{IEEEproof}
  See the Appendix C of \cite{he2017spatial}.
\end{IEEEproof}

Aided with Proposition \ref{proposition2} and the incorporation of a logarithmic barrier function \cite{boyd2004convex}, we therefore seek to solve the following problem:
\begin{equation}
\arraycolsep=1.0pt\def\arraystretch{1.3}
  \begin{array}{cl}
  \displaystyle \max_{\bm\lambda \in \mathbb{R}_{MN_\text{S}\times 1}} & \displaystyle f_\text{B}(\bm\lambda) = R_\text{LB}(\bm\lambda) +  \sum_{i=1}^{MN_\text{S}} \frac{\ln(\lambda_i)}{t_\text{B}} \,\, \text{s.t.} \,\, \mathbf{1}^T \bm\lambda = MN_\text{S} \\
  \end{array},
  \label{P2_1_Barrier}
\end{equation}
where $\lambda_i$ denotes the $i$-th element of $\bm\lambda$, and $t_\text{B}$ is used to scale the penalty of the logarithmic barrier function. The gradient of the cost function in (\ref{P2_1_Barrier}) over $\bm\lambda_m$ is thus given as follows:
\begin{equation}
  \nabla_{\bm\lambda_m}f_\text{B}(\bm\lambda) = \nabla_{\bm\lambda_m} R_\text{LB}(\bm\lambda) + \frac{1}{t_\text{B}} \left[\lambda_{m1}^{-1}, \ldots, \lambda_{mN_\text{S}}^{-1}\right]^T,
  \label{fB_lambda_m}
\end{equation}
where the gradient vector $\nabla_{\bm\lambda_m} R_\text{LB}(\bm\lambda)$ is given in (\ref{RLB_Lambda_m}). Moreover, in order to preserve the linear constraint $\mathbf{1}^T \bm\lambda = MN_\text{S}$, we thus employ the following ascent direction for $\bm\lambda_m$, so that $\mathbf{1}^T \Delta \bm\lambda_m = 0$:
\begin{equation}
  \Delta \bm\lambda_m = \left(\mathbf{I}_{MN_\text{S}} - \frac{\mathbf{1} \cdot \mathbf{1}^T }{M N_\text{S}}\right) \nabla_{\bm\lambda_m} f_\text{B}(\bm\lambda).
\end{equation}

Finally, the ascent direction for $\bm\lambda$ is given as:
\begin{equation}
  \Delta \bm\lambda = \left[\Delta \bm\lambda_1^T, \ldots, \Delta \bm\lambda_{M}^T\right]^T.
  \label{DBF_SearchDirection}
\end{equation}

Using (\ref{DBF_SearchDirection}) as the search direction, the global optimal\footnote{The global optimality is guaranteed due to the concavity of $R_\text{LB}$ over $\bm\lambda$, as verified by Proposition \ref{proposition2}} power allocation vector $\bm\lambda^*$ can thus be obtained via a gradient ascent method. The optimal digital precoder $\mathcal{D}^* = \{\mathbf{D}_1^*, \ldots, \mathbf{D}_M^*\}$ is thus given by:
\begin{equation}
   \mathbf{D}_m^* = \displaystyle \text{diag}\left(\sqrt{\lambda_{m1}^*}, \sqrt{\lambda_{m2}^*}, \ldots, \sqrt{\lambda_{mN_\text{S}}^*}\right),\,\, 1 \le m \le M,
   \label{DBFOutputFormulation}
\end{equation}
where $\bm\lambda^* = [ (\bm\lambda^*_1)^T, \ldots, (\bm\lambda^*_M)^T ]^T$.

\begin{figure*}
    \begin{equation}
      \nabla_\mathbf{a} R_\text{LB}(\mathbf{a}) = \frac{\rho \log_2 e}{M N_\text{S}} \sum_{n=1}^M \frac{\sum_{t=1}^M \left|\bm\Sigma_n + \bm\Sigma_t\right|^{-1} \text{diag}\left[\mathbf{H}^H \left(\bm\Sigma_n + \bm\Sigma_t\right)^{-1} \mathbf{HA} \left(\mathbf{C}_n \bm\Lambda_n \mathbf{C}_n^H + \mathbf{C}_t \bm\Lambda_t \mathbf{C}_t^H \right)\right] }{\sum_{t'=1}^M \left|\bm\Sigma_n + \bm\Sigma_{t'}\right|^{-1}},
      \label{RLB_a}
    \end{equation}
\hrulefill
\end{figure*}

As for (P3), it is found that (P3) is also a non-concave optimization problem, due to the non-convex constraint ($\mathbf{A} \in \mathscr{A}$) and the non-concavity of $R_\text{LB}$ over $\mathbf{A}$. In order to ensure local convergence, we propose to relax (P3) into the following optimization (P3-R) with a convex $\ell_\infty$ constraint:
\begin{equation}
\arraycolsep=1.0pt\def\arraystretch{1.3}
  \begin{array}{lcl}
  \text{(P3-R):    }& \displaystyle \max_{\mathbf{a} \in \mathbb{C}_{N_\text{T}\times 1}}               & \displaystyle R_\text{LB}(\mathbf{a}) \,\, \text{s.t.} \,\, \left\|\mathbf{a}\right\|_\infty \le 1/\sqrt{N_\text{K}},
  \end{array}
  \label{P3_1}
\end{equation}
where $\mathbf{a} = \text{diag}(\mathbf{A}) \in \mathbb{C}_{N_\text{T} \times 1}$ denotes the diagonal elements of $\mathbf{A}$. The $\ell_\infty$ norm is defined as $\|\mathbf{a}\|_\infty = \max_{n=1,\ldots, N_\text{T}}|a_n|$.

In order to deal with the non-differentiable $\ell_\infty$ constraint in (P3-R), similar to \cite{he2015infinity}, we use $\ell_p$ norm to approximate the $\ell_\infty$ norm with a large $p$, which leads to formulating the following optimization with a logarithmic barrier:
\begin{equation}
\arraycolsep=1.0pt\def\arraystretch{1.3}
  \begin{array}{cl}
  \displaystyle \max_{\mathbf{a} \in \mathbb{C}_{N_\text{T} \times 1}}  & \displaystyle g_\text{B}(\mathbf{a}, p) = R_\text{LB}(\mathbf{a}) + \frac{1}{t_\text{B}}\ln\left(\frac{1}{\sqrt{N_\text{K}}} - \left\|\mathbf{a}\right\|_p\right).
  \end{array}
  \label{P3_2}
\end{equation}

To solve (\ref{P3_2}) via a gradient ascent method, we use the gradient as the ascent direction, i.e. $\Delta \mathbf{a} = \nabla_{\mathbf{a}} g_\text{B}(\mathbf{a}, p)$. The gradient $\nabla_{\mathbf{a}} g_\text{B}(\mathbf{a}, p)$ is given as follows:
\begin{equation}
  \nabla_{\mathbf{a}} g_\text{B}(\mathbf{a}, p) = \nabla_\mathbf{a} R_\text{LB}(\mathbf{a}) - \frac{\left\|\mathbf{a}\right\|_p^{1-p} \mathbf{p}_\text{a} }{2 \cdot t_\text{B} \left( N_\text{K}^{-1/2} - \left\|\mathbf{a}\right\|_p \right)},
  \label{gB_a}
\end{equation}
where $\nabla_\mathbf{a} R_\text{LB}(\mathbf{a})$ is given by (\ref{RLB_a}) with $\bm\Lambda_n = \text{diag}(\bm\lambda_n)$, and $\mathbf{p}_\text{a} \in \mathbb{C}_{N_\text{T} \times 1}$ is given as:
\begin{equation}
  \mathbf{p}_\text{a} = \left[ a_1 \cdot \left|a_1\right|^{p-2}, a_2 \cdot \left|a_2\right|^{p-2}, \ldots, a_{N_\text{T}} \cdot \left|a_{N_\text{T}}\right|^{p-2} \right]^T.
\end{equation}

Finally, we present our proposed two-step algorithm as follows, where the digital precoder $\mathcal{D}$ and the analog precoder $\mathbf{A}$ are optimized iteratively.

\textbf{1. Initialization}: Given initial solutions $\mathbf{a}^{(0)}$ and $\bm\lambda^{(0)}$. Set iteration index to $i=0$.

\textbf{2. Digital Precoder Optimization}: Based on $\mathbf{a}^{(i)}$, optimize the digital precoder by solving (\ref{P2_1_Barrier}) with a gradient ascent method, which yields $\bm\lambda^{(i+1)}$.

\textbf{3. Analog Precoder Optimization}: Based on $\bm\lambda^{(i+1)}$, optimize the analog precoder by solving (\ref{P3_2}) with a gradient ascent method, which yields $\mathbf{a}^{(i+1)}$.

\textbf{4. Halt or Update}: Let $i \leftarrow i + 1$. Go to Step 2 until convergence.

We refer the interested readers to \cite{he2017spatial} for more details on the proposed algorithm. Note that, due to the global convergence of solving (\ref{P2_1_Barrier}) and the local convergence of solving (\ref{P3_2}), the proposed algorithm thus ensures local convergence to a feasible hybrid precoder.

Finally, we focus on the optimized selection of the parameters $(N_\text{K}, N_\text{M})$, when the product $N_\text{T} = N_\text{K} N_\text{M}$ is invariant. Qualitatively, increasing $N_\text{K}$ leads to reducing $N_\text{M}$, which therefore reduces the SE gain provided by GenSM, as $M = 2^{\lfloor \log_2 \binom{N_\text{M}}{N_\text{RF}} \rfloor}$. However, increasing $N_\text{K}$ also leads to enlarging the antenna group and thus more array gain can be harnessed by the analog precoder. Hence $(N_\text{K}, N_\text{M})$ is the key to achieve a scalable tradeoff between the multiplexing gain and the array gain. We thus seek to optimize the corresponding parameters for maximizing the \textit{average SE}, i.e.
\begin{equation}
\arraycolsep=1.0pt\def\arraystretch{1.3}
  \begin{array}{rcl}
  && \displaystyle \left(N_\text{K}^*, N_\text{M}^* \right) \\
  &=& \displaystyle \arg\max_{(N_\text{K}, N_\text{M})} E_\mathbf{H}\left\{R_\text{LB}\left[\mathbf{H}, \mathcal{D}^*(\mathbf{H}, N_\text{K}, N_\text{M}), \text{...}\right.\right. \\
  && \displaystyle \left.\left. \mathbf{A}^*(\mathbf{H}, N_\text{K}, N_\text{M}) \right]\right\}, \,\, \text{s.t.} \,\, N_\text{K} N_\text{M} = N_\text{T},
  \end{array}
  \label{ParameterOptimization}
\end{equation}
where $\mathcal{D}^*(\mathbf{H}, N_\text{K}, N_\text{M})$ and $\mathbf{A}^*(\mathbf{H}, N_\text{K}, N_\text{M})$ denote the hybrid precoder designed by the proposed algorithm based on given $\mathbf{H}$, $N_\text{K}$ and $N_\text{M}$. Here we use $R_\text{LB}$ instead of $R$ as the cost function in (\ref{ParameterOptimization}) so that a lower computational complexity can be achieved.

\textit{Remark}: It is worth noting that the conventional hybrid precoding schemes can be conceived as special cases of the proposed framework, when $(N_\text{K}, N_\text{M}) = (N_\text{T}/N_\text{RF}, N_\text{RF})$. Therefore the proposed scheme has the potential to outperform the conventional schemes, given that $(N_\text{K}, N_\text{M})$ are optimized according to (\ref{ParameterOptimization}).

\section{Simulation Results}
In this section we present the simulated SE performance yielded by various schemes. Note that the achievable SE performance of the proposed scheme is given by the \textit{true SE expression} $R(\mathbf{H}, \mathcal{D}, \mathbf{A})$ averaged over $1, 000$ random channel realizations. The simulation parameters are specified in accordance to Table \ref{TABLESimu}, unless mentioned otherwise.

\begin{figure}
\center{\includegraphics[width=0.90\linewidth]{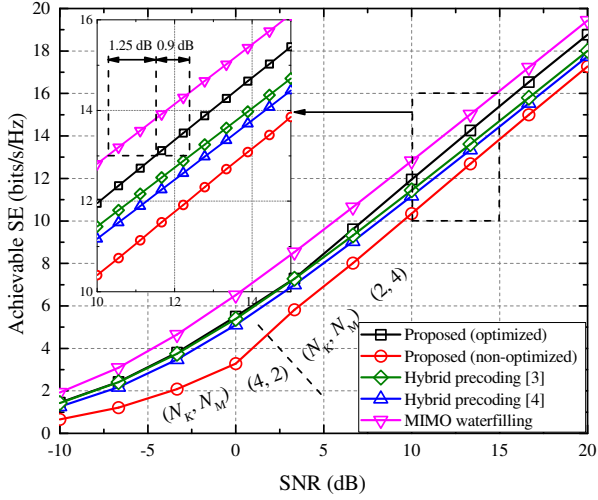}}
\caption{Achievable SE performance yielded by various schemes with $N_\text{T} = 8$, $N_\text{R} = 6$ and $N_\text{RF} = 2$. The parameters $(N_\text{K}, N_\text{M})$ in the proposed GenSM-aided mmWave MIMO scheme is designed by solving (\ref{ParameterOptimization}).}
\label{Fig_Performance8x6}
\end{figure}

In Fig.\ref{Fig_Performance8x6}, the SE performance yielded by various schemes in conjunction with $(N_\text{T}, N_\text{R}, N_\text{RF}) = (8, 6, 2)$ are presented. It can be observed from the figure that $(N_\text{K}, N_\text{M}) = (4, 2)$ and $(N_\text{K}, N_\text{M}) = (2, 4)$ are respectively selected for the proposed scheme (with optimized precoders), when SNR $< 2.5$ dB and SNR $> 2.5$ dB. It is also observed that a significant SE gain is achieved by the proposed scheme with an optimized hybrid precoder compared to the non-optimized counterpart, which substantiates the efficacy of the proposed algorithm. By comparing the proposed scheme with state-of-the-art mmWave schemes, it is seen that our scheme maintains a superior SE performance over \cite{gao2016energy} for the entire SNR range considered, and outperforms the scheme of \cite{ayach2014spatially} when SNR $> 2.5$ dB. Finally, with a target throughput of $13$ bits/s/Hz, our scheme outperforms \cite{ayach2014spatially} by about $0.9$ dB, and is outperformed by the MIMO waterfilling capacity with a $1.25$ dB performance gap.

\begin{figure}
\center{\includegraphics[width=0.90\linewidth]{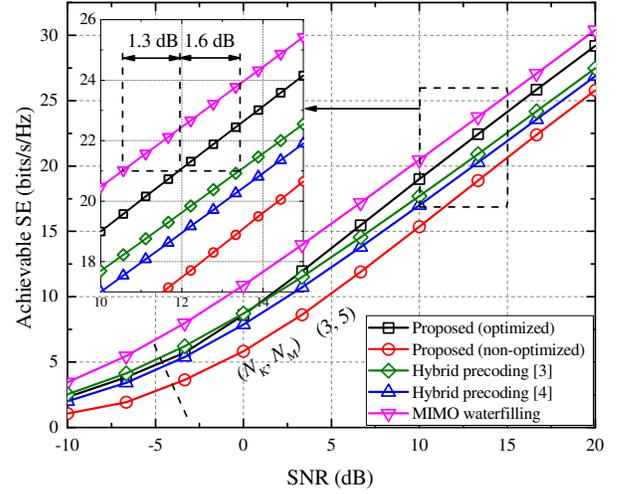}}
\caption{Achievable SE performance yielded by various schemes with $N_\text{T} = 15$, $N_\text{R} = 10$ and $N_\text{RF} = 2$. The parameters $(N_\text{K}, N_\text{M})$ in the proposed GenSM-aided mmWave MIMO scheme is designed by solving (\ref{ParameterOptimization}).}
\label{Fig_Performance15x10}
\end{figure}

In Fig.\ref{Fig_Performance15x10}, we increase the number of RF chains to $N_\text{RF}=3$ in a $15 \times 10$ mmWave MIMO system, and observe that $(N_\text{K}, N_\text{M}) = (5, 3)$ and $(N_\text{K}, N_\text{M}) = (3, 5)$ are respectively selected when SNR $< -7.5$ dB and SNR $> -7.5$ dB. Similar to Fig.\ref{Fig_Performance8x6}, it can be seen that the proposed scheme outperforms \cite{gao2016energy} over the entire SNR range. Finally, with a target throughput of approximately $21$ bits/s/Hz, it is seen that the proposed scheme outperforms \cite{ayach2014spatially} by approximately $1.6$ dB, and is outperformed by the MIMO waterfilling capacity with a $1.3$ dB performance gap.


\section{Conclusion}
In this paper, we proposed a novel GenSM-aided mmWave MIMO scheme with hybrid precoding to improve the achievable SE of conventional mmWave MIMO schemes. A closed-form lower bound was also proposed to provide an accurate approximation to the achievable SE performance. A two-step algorithm was proposed to optimize the hybrid precoder with respect to maximizing the SE lower bound. Finally, numerical simulations are conducted to substantiate the SE gain achieved by the proposed scheme.

\section*{Acknowledgment}
This work was supported by the National Natural Science Foundation of China (Grant No. 61471221 and No. 61471219).

\end{document}